\documentclass[sn-mathphys]{sn-jnl}% Math and Physical Sciences Reference Style
\newcommand{\Jpsi}{$J/\psi$}
\newcommand{\Ups}{$\Upsilon$(1S)}
\usepackage{color}

\usepackage[normalem]{ulem} % \sout{old text} for strikeout
\renewcommand\sout{\bgroup \color[rgb]{0.55,0.00,0.99} \ULdepth=-.5ex \ULset}

\begin{document}

\title[Article Title]{Pion PDFs confronted by Fixed-Target Charmonium
  Production}

\author*[1]{\fnm{Wen-Chen} \sur{Chang}}\email{changwc@phys.sinica.edu.tw}
\author[1]{\fnm{Chia-Yu} \sur{Hsieh}}\email{cyhsieh@phys.sinica.edu.tw}
\author[1]{\fnm{Yu-Shiang} \sur{Lian}}\email{yslian@gate.sinica.edu.tw}
\author[2]{\fnm{Jen-Chieh} \sur{Peng}}\email{jcpeng@illinois.edu}
\author[3]{\fnm{Stephane} \sur{Platchkov}}\email{Stephane.Platchkov@cern.ch}
\author[4]{\fnm{Takahiro} \sur{Sawada}}\email{sawada@icrr.u-tokyo.ac.jp}

\affil[1]{\orgdiv{Institute of Physics}, \orgname{Academia Sinica}, \city{Taipei}, \postcode{11529}, \country{Taiwan}}

\affil[2]{\orgdiv{Department of Physics}, \orgname{University of Illinois at
  Urbana-Champaign}, \city{Urbana}, \postcode{61801}, \country{USA}}

\affil[3]{\orgdiv{IRFU, CEA}, \orgname{Universit\'{e} Paris-Saclay}, \city{Gif-sur-Yvette}, \postcode{91191}, \country{France}}

\affil[4]{\orgdiv{Institute for Cosmic Ray Research}, \orgname{The University of Tokyo}, \city{Gifu}, \postcode{506-1205}, \country{Japan}}

\abstract{The pion, as the Goldstone boson of the strong interaction,
  is the lightest QCD bound state and responsible for the long-range
  nucleon-nucleon interaction inside the nucleus. Our knowledge on the
  pion partonic structure is limited by the existing Drell-Yan data
  which are primarily sensitive to the pion valence-quark
  distributions. The recent progress of global analysis of pion's
  parton distribution functions (PDFs) utilizing various experimental
  approaches are introduced. From comparisons between the pion-induced
  $J/\psi$ and $\psi(2S)$ production data with theoretical
  calculations using the CEM and NRQCD models, we show how these
  charmonium production data could provide useful constraints on the
  pion PDFs.}

\keywords{Pion PDFs, Charmonium, CEM, NRQCD}



\maketitle

%%%%%%%%%%%%%%%%%%%%%%
\section{Introduction}
\label{sec:intro}
%%%%%%%%%%%%%%%%%%%%%%

The pion, being the Goldstone boson of dynamical chiral symmetry
breaking of the strong interaction, is also the lightest QCD bound
state. Because of its light mass, the pion plays a dominant role in
the long-range nucleon-nucleon interaction. Understanding the pion's
internal structure is important to investigate the low-energy,
nonperturbative aspects of QCD~\cite{Horn:2016rip}. Even though the
pion is theoretically simpler than the proton, its partonic structure
is much less explored. As scattering off a pion target is not
feasible, current knowledge on pion PDFs mostly relies on the
pion-induced Drell-Yan data~\cite{Chang:2013opa}. Through the
Drell-Yan reaction, the valence-quark distributions at $x >0.2$ can be
determined while additional measurements are required to constrain the
sea and gluon densities.

While the prompt-photon production process $\pi N \rightarrow \gamma
X$~\cite{WA70:1987bai} was used to constrain the gluon content of
pions through the $Gq \rightarrow \gamma q$ subprocess, the
experimental uncertainties are large. Production of heavy quarkonia,
like \Jpsi~ and \Ups, with a pion beam has distinctive advantages: the
cross sections are large and their decay can be readily detected via
the dimuon decay channel. These datasets have been shown to be
sensitive to both the quark and gluon distributions of the incident
pion~\cite{Gluck:1977zm,Barger:1980mg}. The other interesting approach
of accessing the pion PDFs from the Sullivan
process~\cite{Sullivan:1971kd} in leading neutron deep inelastic
scattering (DIS) data has been considered with promising
results~\cite{Khoze:2006hw, McKenney:2015xis}. This method is subject
to large systematic uncertainties due to the off-shell nature of
virtual pion in the fluctuated Fock state, and further theoretical
studies are required to clarify the uncertainties~\cite{Qin:2017lcd,
  Perry:2018kok}.

In the fixed-target energy domain, where the transverse momentum of
the charmonium \Jpsi~ and $\psi(2S)$ is less than its mass, the
charmonium production is dominated by the quark-antiquark ($q
\bar{q}$) and gluon-gluon fusion ($GG$) partonic processes. The shape
of the longitudinal momentum $x_F$ cross section is sensitive to the
quark and gluon parton distributions of colliding hadrons. Since the
nucleon PDFs are known with good accuracy, the measurement of total as
well as the differential $x_F$ distribution of charmonia with the pion
beam provides, within the theoretical model uncertainties, valuable
information about the pion quark and gluon partonic distributions.

In this article, we present our recent studies about the possibility
to constrain pion gluon density from the existing fixed-target
charmonium data~\cite{Chang:2020rdy, Hsieh:2021yzg, Chang:2022pcb}. We
start with an introduction of various pion PDFs and their distinctive
features in Sec.~\ref{sec:PDFs}, followed by Sec.~\ref{sec:model}
describing the two theoretical frameworks, CEM and NRQCD, used for
describing the charmonium production. Sec.~\ref{sec:results} shows the
comparison of data and theoretical predictions, from which the
differentiation of the large-$x$ gluon strengths in various pion PDFs
can be observed. We conclude with a summary of the results and a few
remarks.

%%%%%%%%%%%%%%%%%%%%%%
\section{Pion PDFs}
\label{sec:PDFs}
%%%%%%%%%%%%%%%%%%%%%%

Pion-induced Drell-Yan data have been included in all global analyses
for the determination of the pion PDFs. However, Drell-Yan process is
mainly sensitive to the valence-quark distribution. Without additional
observables, the sea and gluon distributions can be only inferred
through the momentum and valence-quark sum rules. Different approaches
have been taken to access the gluon and sea quark distributions: (i)
utilizing \Jpsi~ production data in OW~\cite{Owens:1984zj}; (ii)
utilizing the direct-photon production data in
ABFKW~\cite{Aurenche:1989sx}, SMRS~\cite{Sutton:1991ay},
GRV~\cite{Gluck:1991ey}, and xFitter~\cite{Novikov:2020snp}; (iii)
utilizing the leading neutron DIS (LN) in JAM~\cite{Barry:2018ort};
(iv) utilizing the production cross sections at the region of large
transverse momentum ($p_T$) sensitive to NLO $qG$ process in
JAM~\cite{Cao:2021aci}.

In addition, some pion PDFs are constructed based on theoretical
modeling. For example, GRS~\cite{Gluck:1999xe} utilized a constituent
quark model to relate the gluon and antiquark density, and
BS~\cite{Bourrely:2018yck, Bourrely:2020izp, Bourrely:2022mjf} assumed
quantum statistical distributions for all parton species with a
universal temperature. The soft-gluon threshold resummation correction
is known to modify the extraction of valence-quark distribution toward
$x=1$~\cite{Aicher:2010cb} and how this effect modifies the large-$x$
behavior of valence quarks in a global analysis is recently
examined~\cite{Barry:2021osv}. We summarize the data sets used for
various global analyses of pion PDFs in Table~\ref{tab_PDF}.

\begin{table}[!ht]   %\footnotesize
%\setlength\tabcolsep{2pt}
%\addtolength{\tabcolsep}{2pt}
\centering
%\begin{center}
\begin{tabular}{|c|c|c|c|c|c|}
%\hline
\hline
PDFs & DY ($x_F$, $p_T$) & Direct $\gamma$ & $J/\psi$ & LN & Ref. \\
\hline
OW & $\surd$ &  & $\surd$ &   & \cite{Owens:1984zj} \\
ABFKW & $\surd$ &  $\surd$ &  &  & \cite{Aurenche:1989sx} \\
SMRS & $\surd$ &  $\surd$ &  &  & \cite{Sutton:1991ay} \\
GRV & $\surd$ & $\surd$ &  &  & \cite{Gluck:1991ey} \\
GRS & $\surd$ &  &  &   & \cite{Gluck:1999xe} \\
JAM18 & $\surd$ &  &  & $\surd$  & \cite{Barry:2018ort} \\
BS & $\surd$ &  &  &   & \cite{Bourrely:2018yck, Bourrely:2020izp, Bourrely:2022mjf} \\
xFitter & $\surd$ & $\surd$ & &   & \cite{Novikov:2020snp} \\
JAM21 & $\surd$ &  & & $\surd$ & \cite{Cao:2021aci} \\
\hline
\end{tabular}
%\end{center}
\caption {Pion PDFs and utilized data sets.}
\label{tab_PDF}
\end{table}

\begin{figure}[!ht]
\centering
\includegraphics[width=1.0\columnwidth]{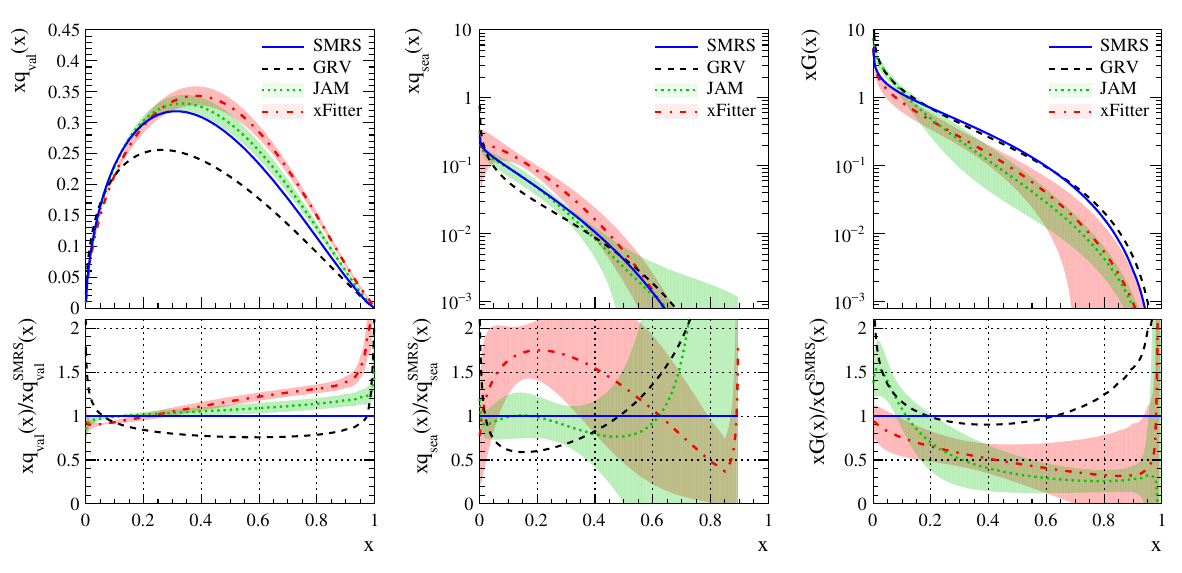}
\caption
[\protect{}] {Momentum density distributions $[xf(x)]$ of valence
  quarks, sea quarks and gluons of SMRS, GRV, xFitter and JAM pion
  PDFs and their ratios to the SMRS PDFs, at the scale of $J/\psi$
  mass ($Q^2$= 9.6 GeV$^2$)~\cite{Chang:2022pcb}. The quark flavor
  ($q$) is either u or d. The uncertainty bands associated with JAM
  and xFitter PDFs are shown.}
\label{fig_PDF}
\end{figure}

Figure~\ref{fig_PDF} compares the valence, sea, and gluon momentum
distributions of the SMRS, GRV, JAM and xFitter pion PDFs at the scale
of \Jpsi~ mass~\cite{Chang:2022pcb}. Their ratios to SMRS are shown in
the bottom panel. Within the range of $x \sim$0.1--0.8, the
valence-quark distributions of SMRS, JAM and xFitter are close to each
other, whereas GRV is lower by up to 20\%--30\%. The sea distribution
shows large variations between the four PDFs. The gluon distributions
also show sizable differences; e.g., in the region of $x > 0.2$ the
xFitter and JAM distributions are smaller in comparison with SMRS and
GRV, by up to a factor of 2-3. As we will see in
Sec.~\ref{sec:results}, these differences in the large-$x$ gluon
distributions lead to quantitative difference in the data description
of fixed-target charmonium data.

%%%%%%%%%%%%%%%%%%%%%%
\section{CEM and NRQCD Models for Charmonium Production}
\label{sec:model}
%%%%%%%%%%%%%%%%%%%%%%

Based on factorization, the theoretical description of charmonium
production consists of the pQCD description of the production of $c
\bar{c}$ pairs at the parton level~\cite{Nason:1987xz, Nason:1989zy,
  Mangano:1992kq}, and their subsequent hadronization into the
charmonium bound state~\cite{Brambilla:2010cs, Lansberg:2019adr}. The
latter nonperturbative part is challenging and has been modeled in
theoretical approaches such as the color evaporation model
(CEM)~\cite{Einhorn:1975ua, Fritzsch:1977ay, Halzen:1977rs}, the
color-singlet model (CSM)~\cite{Chang:1979nn, Berger:1980ni,
  Baier:1983va}, and the nonrelativistic QCD
(NRQCD)~\cite{Bodwin:1994jh, Beneke:1996tk}.

The CEM assumes a constant probability $F^{H}$, specific for each
charmonium $H$, for the hadronization of $c \bar{c}$ pairs into the
colorless hadron state. The differential cross section $d\sigma/dx_F$
for \Jpsi~ from the $\pi N$ collision is expressed as an integration
of $c \bar{c}$ pair production with an invariant mass $M_{c\bar{c}}$
up to the $D \bar{D}$ threshold,
\begin{align}
\frac{d\sigma^{H}}{dx_F}=& F^{H} \sum\limits_{i,j=q, \bar{q}, G} \int_{2 m_c} ^{2 m_{D}} dM_{c \bar{c}} \frac{2M_{c \bar{c}}}{s\sqrt{x_F^2+4{M_{c \bar{c}}}^2/s}} \nonumber \\
 \times f^{\pi}_{i}(x_1, \mu_{F}) & f^{N}_{j}(x_2, \mu_{F}) \hat{\sigma}[ij \rightarrow c \bar{c} X](x_1 p_{\pi} , x_2 p_{N} , \mu_{F}, \mu_{R}),
\label{eq:eq1}
\end{align}
\begin{align}
  x_F = 2 p_L/\sqrt{s} \mbox{,   } x_{1,2} =  \frac{\sqrt{x_F^2+4{M_{c \bar{c}}}^2/s} \pm x_F}{2} 
\end{align}
where $i$ and $j$ denote the interacting partons (gluons, quarks and
antiquarks) and $m_c$, $m_D$, and $M_{c \bar{c}}$ are the masses of
the charm quark, $D$ meson, and $c \bar{c}$ pair, respectively. The
$f^{\pi}$ and $f^{N}$ are the corresponding pion and nucleon parton
distribution functions, respectively, evaluated at the corresponding
Bjorken-$x$, $x_1$ and $x_2$, at the factorization scale $\mu_F$. The
short-distance differential cross section of heavy-quark pair
production $\hat{\sigma}[ij \rightarrow c \bar{c} X]$ is calculable as
a perturbation series in the strong coupling $\alpha_s(\mu_R)$
evaluated at the renormalization scale $\mu_R$. The longitudinal
momentum of the experimentally detected dilepton pair, equivalent to
that of the $c \bar{c}$ pair, is denoted by $p_L$.

The $F^H$ factor is to be determined as the normalization parameter in
the fit to the experimental measurements. The assumption of a common
$F^H$ factor for different subprocesses greatly reduces the number of
free parameters of the CEM. In spite of its well-known
limitations~\cite{Bodwin:2005hm}, the CEM gives a good account of many
features of fixed-target \Jpsi~ cross section data with proton beams,
including their longitudinal momentum ($x_F$)
distributions~\cite{Gavai:1994in, Schuler:1996ku} and the collider
data at RHIC, Tevatron, and LHC~\cite{Nelson:2012bc,
  Lansberg:2020rft}.

\begin{table}[!ht]   %\footnotesize
%\setlength\tabcolsep{2pt}
%\addtolength{\tabcolsep}{2pt}
\centering
%\begin{center}
\begin{tabular}{|c|c|c|c|}
%\hline
\hline
 $H$ & $q \bar{q}$ & $GG$ & $qG$ \\
\hline
$J/\psi$, $\psi(2S)$ & $\langle \mathcal{O}_{8}^{H}[^{3}S_{1}] \rangle$ ($\alpha_{s}^2$) & $ \Delta_8^{H}$ ($\alpha_{s}^2$) & \\
 & &  $\langle \mathcal{O}_{1}^{H}[^{3}S_{1}] \rangle$ ($\alpha_{s}^3$) & \\
\hline
$\chi_{c0}$ & $\langle \mathcal{O}_{8}^{H}[^{3}S_{1}] \rangle$ ($\alpha_{s}^2$) & $ \langle \mathcal{O}_{1}^{H}[^{3}P_{0}] \rangle$ ($\alpha_{s}^2$) & \\
\hline
$\chi_{c1}$ & $\langle \mathcal{O}_{8}^{H}[^{3}S_{1}] \rangle$ ($\alpha_{s}^2$) & $ \langle \mathcal{O}_{1}^{H}[^{3}P_{1}] \rangle$ ($\alpha_{s}^3$) & $ \langle \mathcal{O}_{1}^{H}[^{3}P_{1}] \rangle$ ($\alpha_{s}^3$)\\
\hline
$\chi_{c2}$ & $\langle \mathcal{O}_{8}^{H}[^{3}S_{1}] \rangle$ ($\alpha_{s}^2$) & $ \langle \mathcal{O}_{1}^{H}[^{3}P_{2}] \rangle$ ($\alpha_{s}^2$) & \\
\hline
\end{tabular}
%\end{center}
\caption {Relationship of LDMEs and the associated orders of
  $\alpha_s$ to the scattering subprocesses for various charmonium states
  in the NRQCD framework of Ref.~\cite{Beneke:1996tk}.  Here
  $\Delta_8^{H} = \langle \mathcal{O}_{8}^{H}[^{1}S_{0}] \rangle +
  \frac{3}{m_c^2} \langle \mathcal{O}_{8}^{H}[^{3}P_{0}] \rangle +
  \frac{4}{5m_c^2} \langle \mathcal{O}_{8}^{H}[^{3}P_{2}] \rangle $.}
\label{tab:LDMEproc}
\end{table}

To examine a possible model dependence of observations, we carry out a
similar study using NRQCD. The NRQCD factorization formula allows for
a systematic expansion of inclusive quarkonium cross sections in
powers of the strong coupling constant $\alpha_s$ and the relative
velocity $v$ of the heavy quarks. This expansion takes into account
the short-distance production of color-singlet and color-octet
$c\bar{c}$ precursor states with various spin ($S$), color ($n$), and
angular momentum ($J$) quantum numbers. The long-distance matrix
elements (LDMEs) are non-perturbative parameters that characterize the
probability of a $c\bar{c}$ pair to evolve into a final quarkonium
state. The LDMEs, assumed to be universal, are extracted from the
experimental data. The differential cross section $d\sigma/dx_F$ for
\Jpsi~ from the $\pi N$ collision is expressed as follows,
\begin{align}
\label{eq:eq2}
\frac{d\sigma^{H}}{dx_F}=& \sum\limits_{i,j=q, \bar{q}, G} \int_{0} ^{1} dx_{1} dx_{2} \delta(x_F - x_1 + x_2) \nonumber \\
 \times& f^{h}_{i}(x_1, \mu_{F}) f^{N}_{j}(x_2, \mu_{F}) \hat{\sigma}[ij \rightarrow H](x_1 P_{h} , x_2 P_{N} , \mu_{F}, \mu_{R}, m_c), \\
\hat{\sigma}[ij \rightarrow H] =&  \sum\limits_{n} \hat{\sigma}[ij \rightarrow c \bar{c} [n]] (x_1 P_{h} , x_2 P_{N} , \mu_{F},
\mu_{R}, m_c) \langle \mathcal{O}_{n}^{H}[^{2S+1}L_{J}] \rangle
\end{align}
where $\hat{\sigma}[ij \rightarrow c \bar{c} [n]]$ denotes the
hard-QCD production cross section for $c \bar c$ pair of color state
$n$ and $\langle \mathcal{O}_{n}^{H}[^{2S+1}L_{J}] \rangle$ is the
corresponding LDME. Table~\ref{tab:LDMEproc} summarizes the
relationships between the LDMEs and the scattering subprocesses for
$J/\psi$, $\psi(2S)$, $\chi_{c0}$, $\chi_{c1}$, and $\chi_{c2}$, up to
$\mathcal{O}(\alpha_{s}^3)$ in the NRQCD
framework~\cite{Beneke:1996tk} adopted for computing $J/\psi$,
$\psi(2S)$, and $\chi_{cJ}$ production via $GG$, $q \bar{q}$ and $qG$
subprocesses. The $J/\psi$ cross section is estimated taking into
account the direct production of $J/\psi$ and the feed-down from
hadronic decays of $\psi(2S)$ and radiative decays of three
$\chi_{cJ}$ states.

%%%%%%%%%%%%%%%%%%%%%%
\section{Results and Discussions}
\label{sec:results}
%%%%%%%%%%%%%%%%%%%%%%

\subsection{Integrated cross sections}

\begin{figure}[!ht]
\centering
\includegraphics[width=0.8\columnwidth]{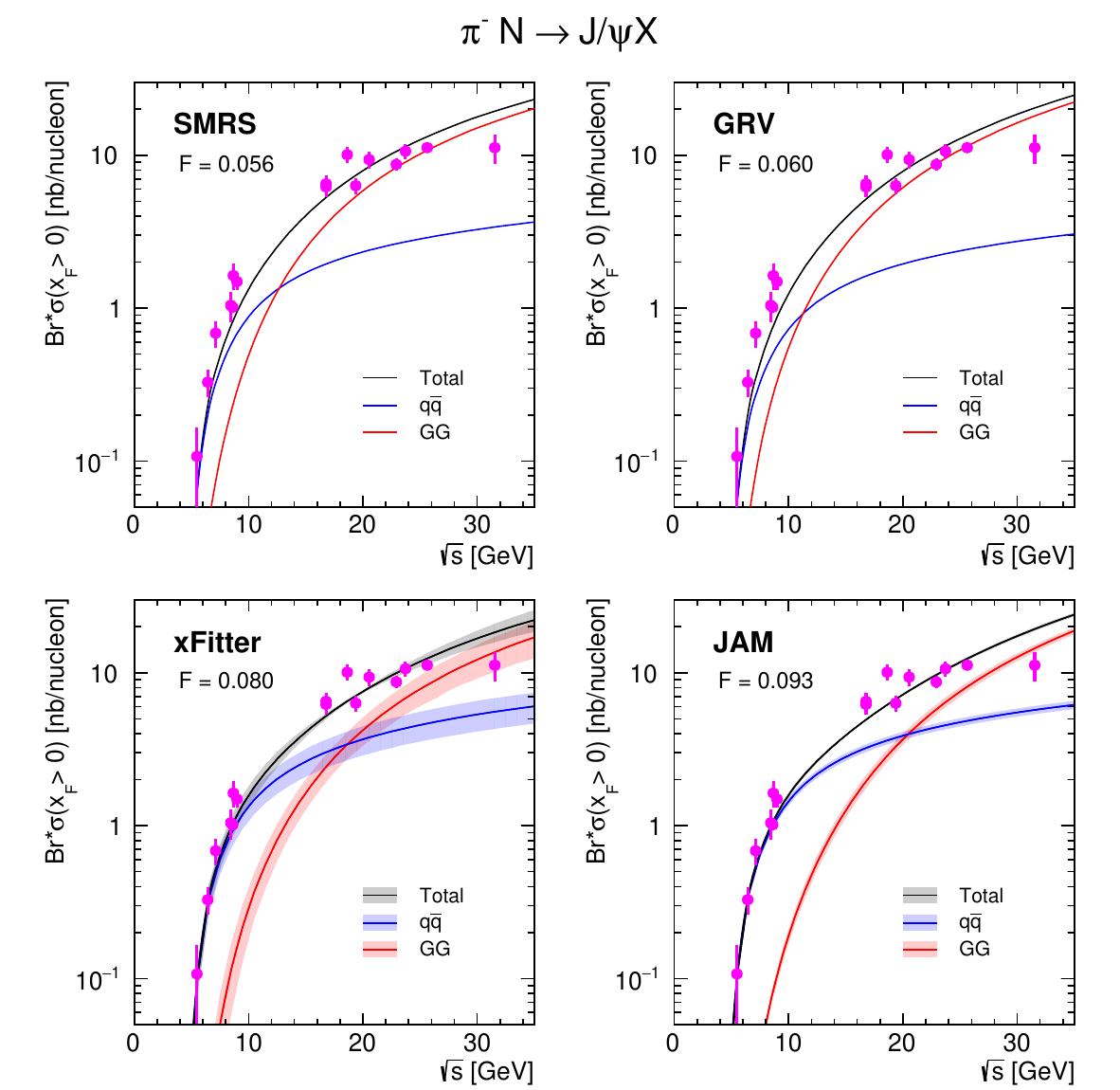}
\caption[\protect{}]{Comparison of \Jpsi~ dimuon decay branching ratio
  ($\rm{Br}$) and \Jpsi~ production cross sections at $x_F >0$ for the
  $\pi^- N$ reaction, calculated by the NLO CEM with four pion PDFs
  (SMRS, GRV, xFitter and JAM) with the data (solid
  circles~\cite{Schuler:1994hy,
    BEATRICE:1999mqh})~\cite{Chang:2020rdy}. The black, blue, and red
  curves represent the calculated total cross section and the $q
  \bar{q}$ and $GG$ contributions, respectively. The shaded bands on
  the xFitter and JAM calculations represent the uncertainties of the
  corresponding PDF sets.}
\label{fig_jpsi_sdep_CEM}
\end{figure}

\begin{figure}[!ht]
\centering
\includegraphics[width=0.8\columnwidth]{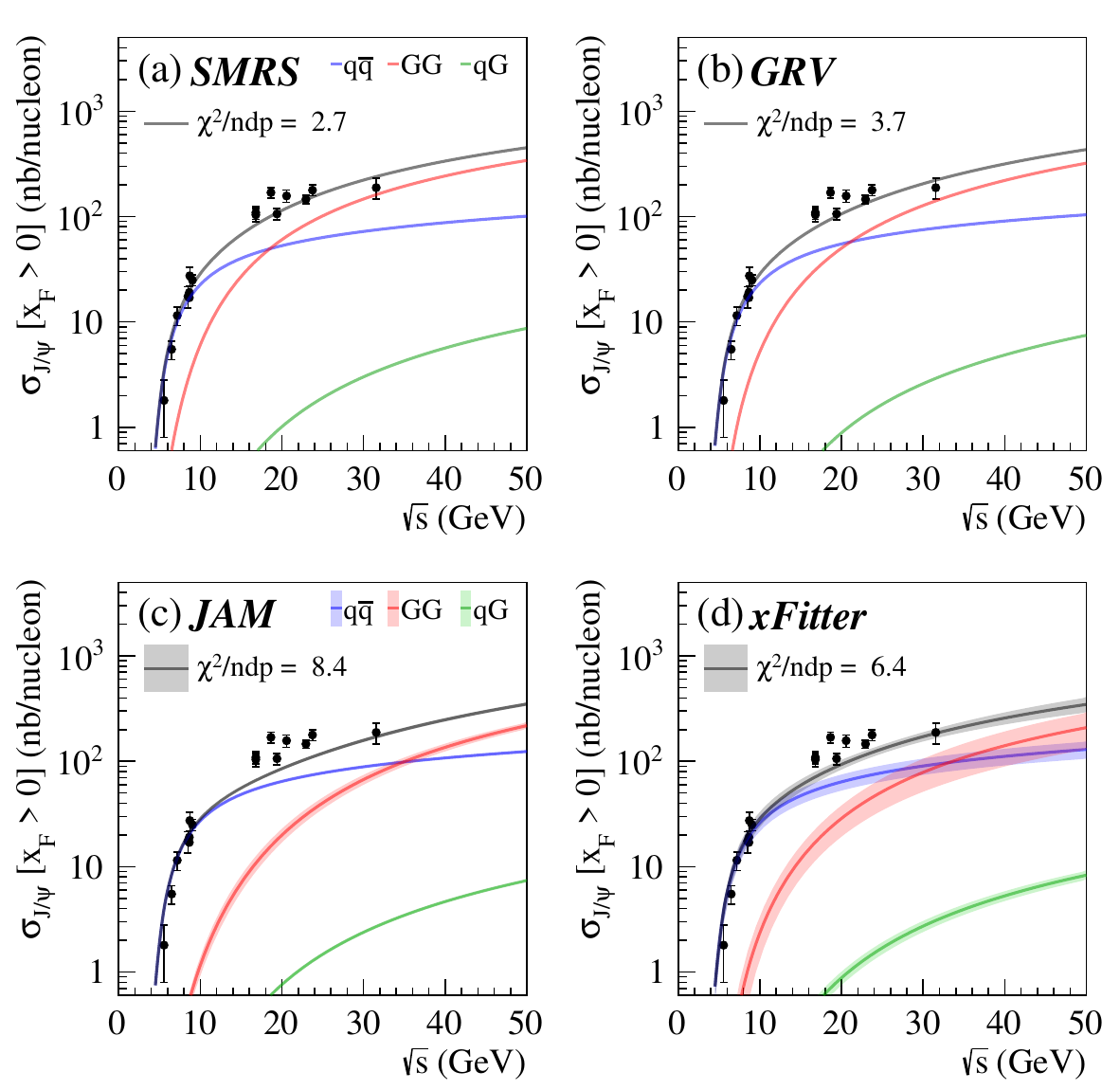}
\caption
[\protect{}] {Same as Fig.~\ref{fig_jpsi_sdep_CEM} with the NRQCD
  calculations ~\cite{Hsieh:2021yzg}.}
\label{fig_jpsi_sdep_NRQCD}
\end{figure}

We start with the comparison between the data of $\pi^- N \rightarrow
J/\psi X$ cross sections integrated over $x_F >
0$~\cite{Schuler:1994hy, BEATRICE:1999mqh} and the NLO CEM
calculations with four pion PDFs, shown in
Fig.~\ref{fig_jpsi_sdep_CEM}. The evaluation of cross sections is done
with a charm quark mass $m_c = 1.5$ GeV/$c^2$ and renormalization and
factorization scales of $\mu_R = m_c$ and $\mu_F = 2 m_c$,
respectively. The hadronization factors $F$ in the CEM model are
assumed to be energy independent and determined by the best fit to the
data for the central values of each pion PDF. The differences between
them are visible through the $F$ factors, which vary from 0.05 to
0.09. Similar comparison made for the NRQCD calculations is shown in
Fig.~\ref{fig_jpsi_sdep_NRQCD}.

In the CEM study, the factor $F$ is determined by the best $\chi^2$
fit to each data set individually. In contrast, a global analysis of
all data sets was performed to obtain some color-octet LDMEs as the
fit parameters in the study with NRQCD. The quality of data
description for each data set in NRQCD study is shown by
$\chi^2/\text{ndp}$, where ndp denotes the number of degree of data
points in a specific data set.

The total cross sections evaluated with the four PDFs exhibit quite
similar $\sqrt{s}$ dependencies, and all agree reasonably with the
data. The $q \bar{q}$ contribution dominates at low energies, whereas
the $GG$ contribution becomes important with increasing
$\sqrt{s}$. The relative fractions of $q \bar{q}$ and $GG$
contributions as a function of $\sqrt{s}$ vary for each pion PDFs,
reflecting the differences between the corresponding parton
distributions. For SMRS and GRV the $GG$ contribution starts to
dominate the cross section around $\sqrt{s}=15$ GeV. For xFitter and
JAM the corresponding values are larger at $\sim$$\sqrt{s}=20-30$ GeV
because of their relatively reduced gluon strength in the valence
region.

\subsection{Differential $x_F$ cross sections}

\begin{figure}[!ht]
\centering
\includegraphics[width=0.8\columnwidth]{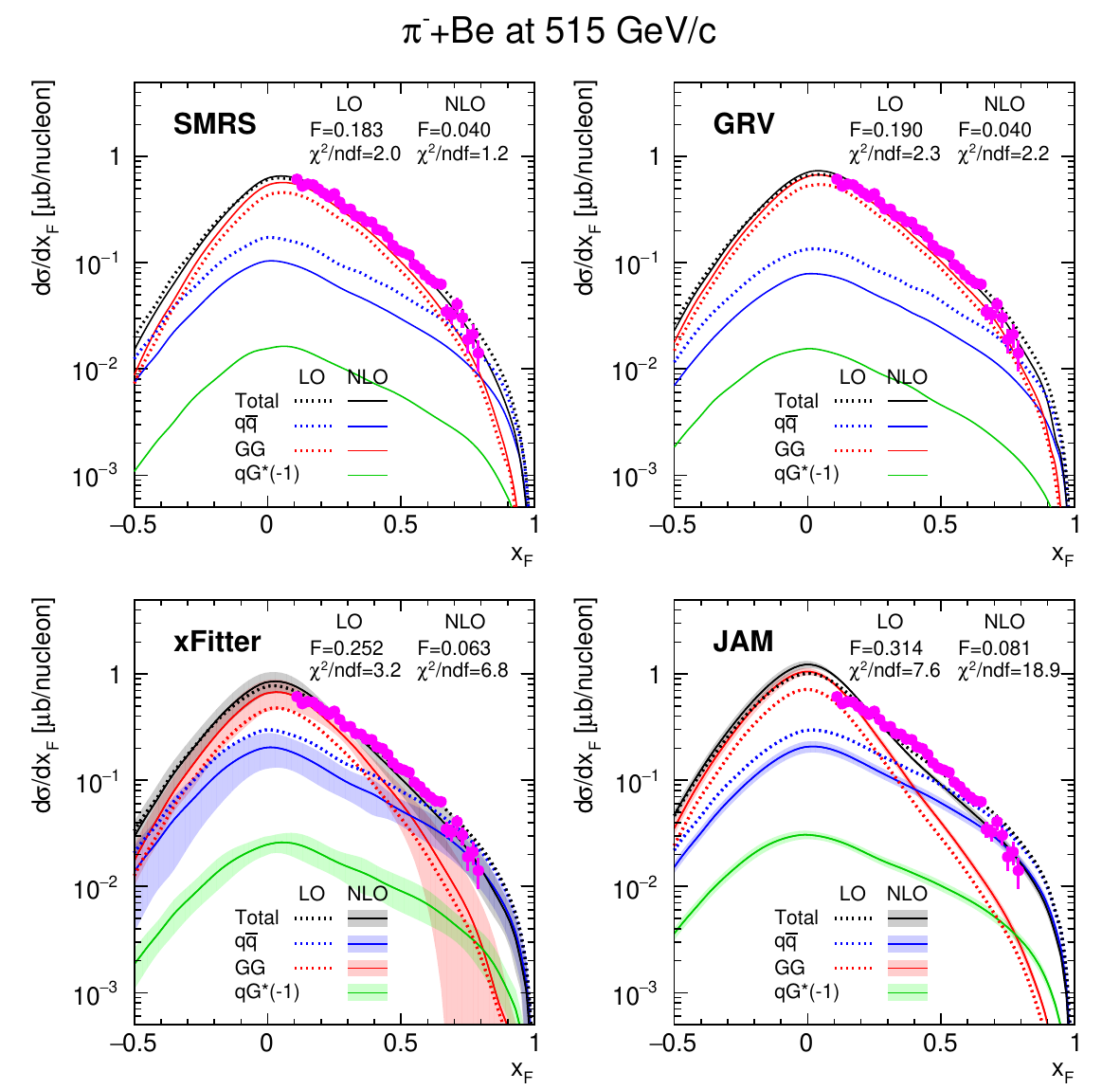}
\caption[\protect{}]{Comparison of the LO and NLO CEM results for the
  SMRS, GRV, xFitter, and JAM PDFs, with the $d\sigma/dx_F$
  data~\cite{jpsi_data1} of \Jpsi~ production off the beryllium target
  with a 515-GeV/$c$ $\pi^-$ beam~\cite{Chang:2020rdy}. The total
  cross sections and $q \bar{q}$, $GG$, and $qG\times(-1)$
  contributions are denoted as black, blue, red, and green lines,
  respectively. Solid and dotted lines are for the NLO and LO
  calculations, respectively. The shaded bands on the xFitter and JAM
  calculations come from the uncertainties of the corresponding PDF
  sets. The resulting $\chi^2$/ndf and $F$ factors are displayed.}
\label{fig_jpsi_data1_CEM}
\end{figure}

\begin{figure}[!ht]
\centering
\includegraphics[width=0.8\columnwidth]{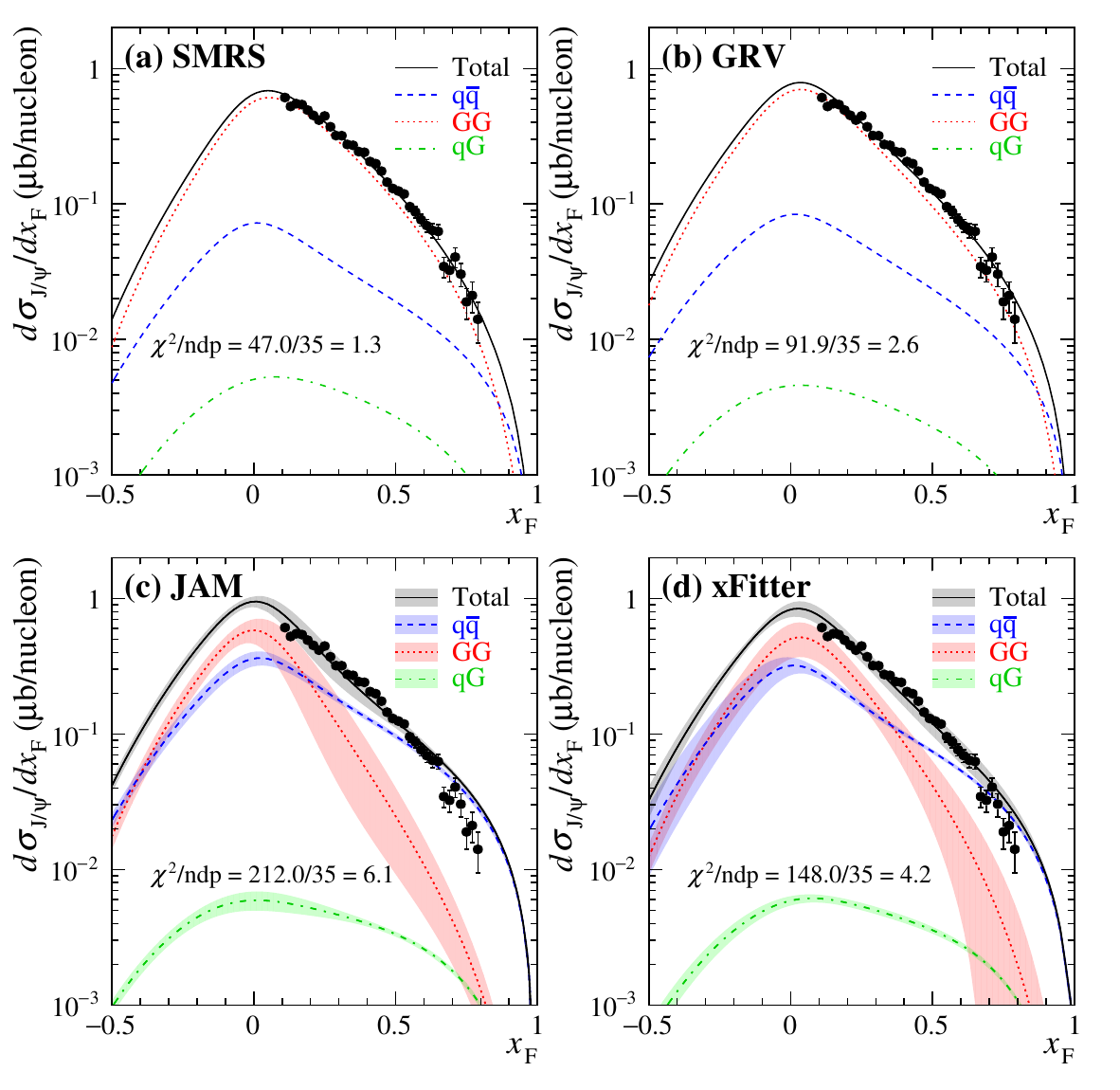}
\caption[\protect{}]{Same as Fig.~\ref{fig_jpsi_data1_CEM} with the
  NRQCD calculations~\cite{Chang:2022pcb}.}
\label{fig_jpsi_data1_NRQCD}
\end{figure}

To investigate further the effect led by different pion PDFs, we
compare the longitudinal $x_F$ distribution of the calculated
pion-induced \Jpsi~ production cross section with a selection of
fixed-target data from Fermilab and CERN experiments for pion-induced
\Jpsi~ production as seen in Table II of Refs.~\cite{Chang:2020rdy,
  Chang:2022pcb}. The beam momenta of the datasets cover the range of
39.5--515 GeV/$c$, corresponding to $\sqrt{s}$ values ranging from 8.6
to 31.1 GeV.

The comparison of our LO and NLO CEM calculations to the E672/E706
data~\cite{jpsi_data1} with a 515 GeV/$c$ $\pi^-$ beam scattered off
Be targets is shown in Fig.~\ref{fig_jpsi_data1_CEM}. Judging from the
reduced $\chi^2$/ndf values, the NLO calculations with SMRS and GRV
are in better agreement with the data than those with xFitter and
JAM. The NLO calculation improves the description of the E672/E706
data only in the cases of SMRS and
GRV. Fig.~\ref{fig_jpsi_data1_NRQCD} shows the same comparison with
the NRQCD caluclations. It is also observed that SMRS and GRV are
favored over JAM and xFitter in both comparisons with the CEM and
NRQCD results.

The fraction of the $GG$ component is maximized around $x_F = 0$,
corresponding to the gluon distribution $G_{\pi}(x)$ around $x
\sim$0.1--0.2. As a result of the rapid drop of the $G_{\pi}(x)$
toward $x=1$, the $GG$ contribution quickly decreases at large
$x_F$. In contrast, the $q \bar{q}$ contribution has a slower fall-off
toward high $x_F$ because of a relatively strong pion valence
antiquark density, in comparison with the gluon one, at large $x$. The
ratio of $q \bar{q}$ to $GG$ shows a strong $x_F$ dependence, making
the $x_F$-differential cross sections at high energies particularly
sensitive to the shape of pion $G_{\pi}(x)$.

%$J/\psi$ and $\psi(2S)$ Ratios

\begin{figure}[!ht]
\centering
\includegraphics[width=0.8\columnwidth]{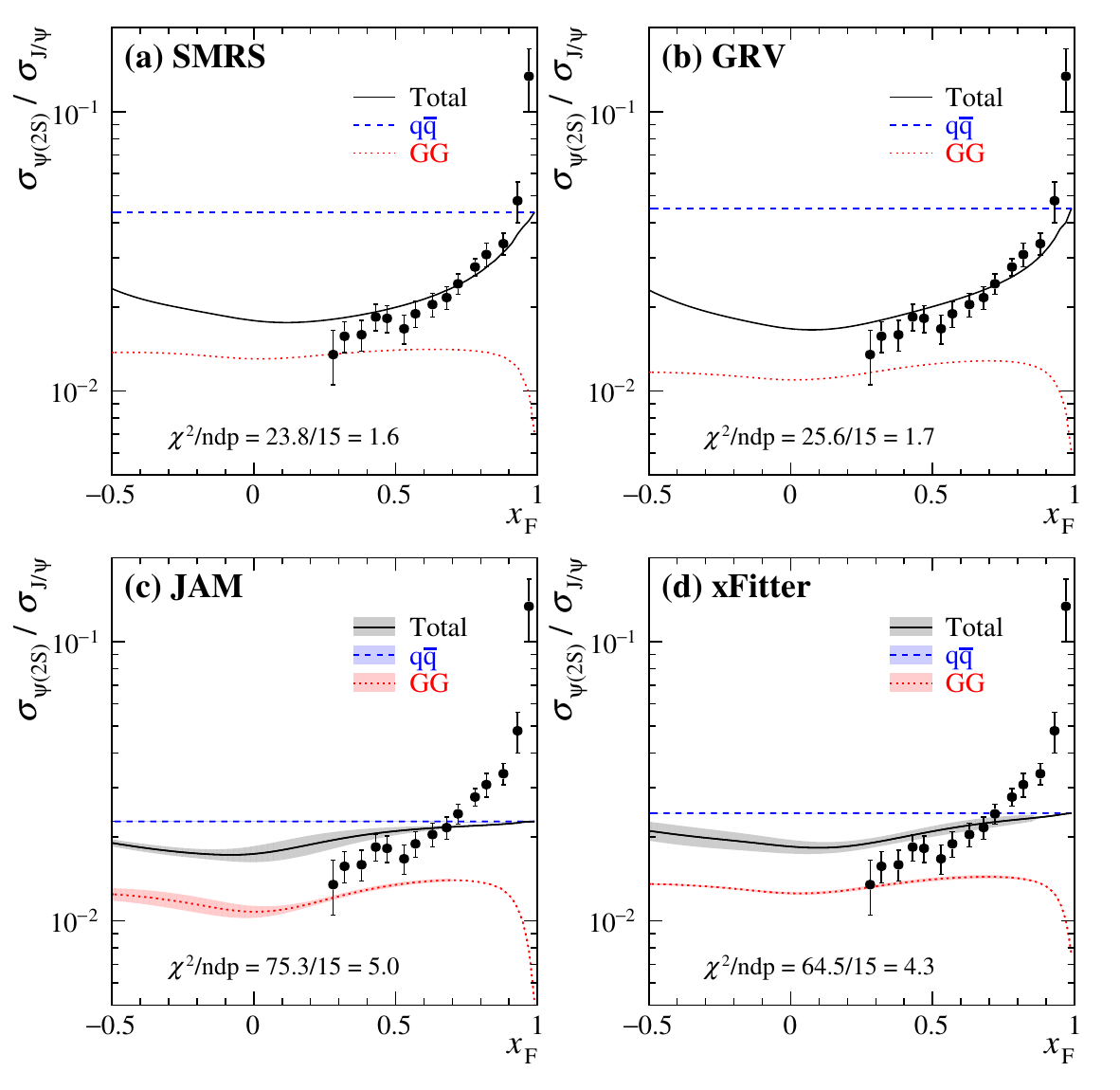}
\caption[\protect{}]{The $\psi(2S)$ to $J/\psi$ cross section ratios
  $R_{\psi}(x_F)$ for $J/\psi$ and $\psi(2S)$ production with a
  252-GeV/$c$ $\pi^-$ beam~\cite{Heinrich:1991zm}. The data are
  compared to the NRQCD calculations for the SMRS, GRV, xFitter, and
  JAM PDFs~\cite{Chang:2022pcb}. The ratios of total cross sections
  and individual $R^{q \bar q}_\psi (x_F)$ and $R^{GG}_\psi (x_F)$
  contributions are denoted as solid black, dashed blue, and dotted
  red lines, respectively.}
\label{fig_jpsi_data10_NRQCD}
\end{figure}

More information on the charmonium production mechanism can be
obtained by comparing the production of the two charmonium states,
$J/\psi$ and $\psi(2S)$. Fig.~\ref{fig_jpsi_data10_NRQCD} shows the
comparison of the $\psi(2S)$ to $J/\psi$ ratios, $R_{\psi}(x_F)$, with
the pion beam momentum of 252 GeV/$c$~\cite{Heinrich:1991zm} and the
NRQCD calculations. An $x_F$-independent $R_{\psi}(x_F)$ is predicted
by the CEM~\cite{HERA-B:2006bhy}, since the fractions of $q \bar{q}$
and $GG$ components are identical for $J/\psi$ and $\psi(2S)$. In
NRQCD, an $x_F$-dependent $R_{\psi}(x_F)$ is possible because
different LDMEs are associated with the $q \bar{q}$ and $GG$ channels
in evaluating the production of $J/\psi$ and $\psi(2S)$.

Fig.~\ref{fig_jpsi_data10_NRQCD} shows a strong $x_F$ dependence of
$R_{\psi}$ and this suggests that the relative weights of the
individual subprocesses $q \bar{q}$ and $GG$ components in $J/\psi$
and $\psi(2S)$ production are distinctly different. The pronounced
rise in the $R_\psi(x_F)$ data at forward $x_F$ where the $ q \bar q$
subprocess dominates the production, indicates that the $ q \bar q$
subprocess is more important for the $\psi(2S)$ production than for
the $J/\psi$ production. The comparison of this result remains to
favor the calculations with SMRS and GRV, consistently with the
observation with $J/\psi$ production data.

%%%%%%%%%%%%%%%%%%%%%%
\section{Summary}
\label{sec:summary}
%%%%%%%%%%%%%%%%%%%%%%

We examine the existing pion PDFs which exhibit pronounced
differences, particularly in their gluon distributions. Using these PDFs
as the input of CEM and NRQCD, the total and $x_F$ differential cross
sections of pion-induced $J/\psi$ and $\psi(2S)$ production are
calculated and compared to the fixed-target data.

We observe the importance of the gluon-gluon fusion process in
charmonium production, especially at high (fixed-target)
energies. Since the calculated shapes of $x_F$ distributions of $GG$
and $q \bar{q}$ contributions are directly related to the parton $x$
distributions of corresponding PDFs, a proper description of
charmonium production data, especially for $x_F>0.5$, imposes strong
constraints on the relevant pion's parton densities. Among the four
pion PDFs examined, both CEM and NRQCD calculations clearly favor SMRS
and GRV PDFs whose gluon densities at $x > 0.1$ are stronger, compared
with xFitter and JAM PDFs. The $GG$ contribution from the latter two
pion PDFs drops too fast toward $x_F = 1$ to describe the data. While
future theoretical developments are required to reduce the theoretical
uncertainties in describing the charmonium production and thus improve
the precision of the extracted PDFs, we emphasize the importance of
including the pion-induced charmonium data in future pion PDF global
analysis.

In the near future, new measurements of Drell-Yan as well as $J/\psi$
data in $\pi^- N$ reactions will be available from the CERN
COMPASS~\cite{COMPASS:2017jbv} and AMBER~\cite{Adams:2018pwt}
experiments. For the coming electron-ion collider projects in U.S. and
China, the pion as well kaon structures are to be explored using the
tagged DIS process~\cite{Aguilar:2019teb, Arrington:2021biu,
  Xie:2021ypc, Chavez:2021koz}. To characterize the recoiled baryon
system from the collisions with very small four-momentum transfer for
the extraction of on-shell meson PDFs, a high-resolution zero-degree
calorimetor is required. A collaboration of East Asian countries on
developing this key detector for U.S. EIC project was recently
discussed~\cite{NCUEIC}.

%%===========================================================================================%%
\backmatter

\bmhead{Acknowledgments}

We thank Nobuo Sato and Ivan Novikov for helping us with the usage of
JAM and xFitter PDFs.

\bmhead{Authors' contributions}

All authors equally contributed to the manuscript. Wen-Chen Chang is
the lead author in organizing and writing the draft. All authors read,
polished, and approved the final manuscript.

\bmhead{Funding} 

This work is supported in part by the U.S. National Science Foundation
and National Science and Technology Council of Taiwan (R.O.C.).

\bmhead{Availability of data and materials}

The datasets generated during and/or analysed during the current study
are available from the corresponding author on reasonable request.

\section*{Declarations}

\bmhead{Ethics approval and consent to participate}

N/A

\bmhead{Consent for publication}

N/A

\bmhead{Competing interests}

The authors declare that they have no competing interests.

\bibliography{ref}% common bib file



\end{document}